\documentclass{article}

\usepackage[english]{babel}

\usepackage[letterpaper,top=2cm,bottom=2cm,left=3cm,right=3cm,marginparwidth=1.75cm]{geometry}
\usepackage{pgfplots}
\usepackage{type1cm}   
\usepackage{booktabs}
\usepackage{makeidx}         
\usepackage{graphicx}        
\usepackage{multicol}        
\usepackage[bottom]{footmisc}

\usepackage{comment}

\usepackage{newtxtext}       %
\usepackage{bm}
\usepackage[varvw]{newtxmath}       
\usepackage{hyperref}
\usepackage{subfig}

\makeindex             

\begin{document}

\title{On the choice of proper outlet boundary conditions for numerical simulation of cardiovascular flows}
\author{Zahra Mirzaiyan$^\star$, 
Michele Girfoglio$^\circ$ 
 and Gianluigi Rozza$^\star$} 
%
%
\maketitle

$^\star$ SISSA, Mathematics Area, mathLab, via Bonomea 265, I-34136 Trieste, Italy, zmirzaiy@sissa.it

$^\circ$ Department of Engineering, University of Palermo, Via delle Scienze, Ed. 7, Palermo, 90128, Italy, michele.girfoglio@unipa.it

$^\star$ SISSA, Mathematics Area, mathLab, via Bonomea 265, I-34136 Trieste, Italy, grozza@sissa.it


\abstract{It is well known that in the computational fluid dynamics  simulations related to the cardiovascular system the enforcement of outflow boundary conditions is a crucial point. In fact, they highly affect the computed flow and a wrong setup could lead to unphysical results. 
In this chapter we discuss the main features of two different ways for the estimation of proper outlet boundary conditions in the context of hemodynamics simulations: on one side, a lumped parameter model of the downstream circulation and, on the other one, a technique based on optimal control.}

\section{Introduction}
\label{sec:1}
The numerical simulation of the Navier-Stokes equations (NSE), describing the dynamics of blood flow in the cardiovascular system, is a fundamental tool when patient-specific configurations are investigated. 
 Since it is prohibitive to perform a discretization of the entire cardiovascular system, NSE 
 are usually solved only on a portion of it, while the missing part is modeled by means of proper
 boundary conditions 
 related to the physiological conditions at the inlets
 and outlets of the domain of interest \cite{Siena:2024}. 
 The choice of proper boundary conditions is a very delicate point in order to get clinically relevant  outcomes: in fact, several studies have demonstrated how this aspect significantly affects the dominant patterns of the computed flow field: see, e.g., \cite{Morbiducci:2010,Van:2011}. 
 Regarding inlet boundary conditions, in most cases, even when patient-specific in vivo data are missing, the enforcement of a generic velocity profile with a pulsatile waveform 
 taken from the
 literature is sufficient to perform reliable numerical simulations \cite{Pedley:1980}. On the contrary, for what concerns outlet boundary conditions, the selection process is more complex and is strictly connected to the availability of clinical data related to the patient at hand. 
 
 In this chapter we go through two different strategies for the setting of proper outlet boundary conditions for cardiovascular applications:
 
 \begin{itemize}
 \item The former 
is related to the employment of zero-dimensional (lumped
 parameter) model \cite{Sankaran:2012,Grinberg:2008} 
  prescribing a specific pressure-flow relationship
 at each outlet of the domain. 
 Lumped models may only include resistances or also involve other elements, such as compliances and inductances.  
 In this framework patient-specific in-vivo measurements can be used to tune the numerical values of the lumped elements. 
 \item  The latter is related to a method based on the solution of an optimal control problem 
 where the mismatch between the numerical solution of NSE 
 and the clinical data is treated as a cost functional to be minimized. 
Here the control variable is the unknown shear stress imposed at each outlet of the domain \cite{Zakia:2021,Fevola:2021}. 
\end{itemize}

We highlight that the two approaches could be also merged as recently has been proposed: for instance, we could consider a zero-dimensional model involving only resistances and then employ an optimal control problem where the control variables are right the unknown resistances: see \cite{Fevola:2021}. However, for sake of clearness, we will treat separately the two strategies.

The rest of the chapter is organized as follows: Section \ref{sec:statement} presents the methodological details of the two approaches mentioned above. Results on two patient-specific cases are reported in Section \ref{numerical_results}, the one where a lumped element model is employed and the other one where the optimal control approach is used. Finally, 
 Section \ref{conclusion}
 provides some concluding remarks. 

\section{Problem statement} 
\label{sec:statement}
We consider the motion of the blood in a time-independent domain $\Omega \subseteq \mathbb{R}^3$ 
over a time interval of interest $(0, T]$. We assume that the blood behaves as a Newtonian fluid. Then the flow is described by unsteady and incompressible NSE:
\begin{equation}
	\begin{cases} 
     \partial_t \bm v
- \nu \Delta \bm{v} + (\bm{v} \cdot \nabla) \bm{v} + \nabla  p  = 0  & \quad \text{in} \quad \Omega \times (0, T], \\
	\nabla \cdot \bm{v} = 0 & \quad \text{in} \quad \Omega \times (0, T],
	\end{cases} \label{tN-Steady}
\end{equation}
where $\partial_t$ denotes the time derivative, $\bm v$ the velocity field, $p$ the blood pressure and $\nu$ the kinematic viscosity.  

Let $\Gamma_i$ be the inlet boundary, $\Gamma_w$ the wall boundary, and $\Gamma_o$ the outlet boundary, such that $\Gamma_i \cup \Gamma_w \cup \Gamma_o = \partial \Omega$ and $\Gamma_i \cap \Gamma_w \cap \Gamma_o = \emptyset$. Then problem \eqref{tN-Steady} is endowed with the following boundary conditions:
\begin{equation}
    \begin{cases}
     \bm v = \bm v_{i}  & \quad \text{on} \quad \Gamma_{i} \times (0, T], \\
     -\nu(\nabla\bm v )\bm n_o + p\bm n_o = \bm g   & \quad \text{on} \quad \Gamma_{o} \times (0, T], \\
     \bm v = \bm 0  & \quad \text{on} \quad \Gamma_{w} \times (0, T],
    \end{cases} \label{tN-Steady-bc}
\end{equation}
and initial data $\bm v = \bm v^0$. In \eqref{tN-Steady-bc} $\bm v_i$ is a given inlet velocity profile, $\bm n_o$ is the outward unit normal vector to $\Gamma_o$ and $\bm g$ is the unknown normal stress modeling the downstream circulation of the patient at hand.

In order to characterize the flow regime under consideration, we define the Reynolds number as
\begin{equation}
Re = \dfrac{UL}{\nu},
\end{equation}\label{eq:Re}
where $U$ and $L$ are characteristic macroscopic velocity and length, respectively. For an internal flow in a cylindrical pipe, $U$ is the
 mean sectional velocity and $L$ is the diameter. 

In order to properly set $\bm g$ in the second equation of \eqref{tN-Steady-bc} we consider two strategies: the employment of a lumped parameter model and of an optimal control approach. 


\subsection{Lumped parameter model}
This approach consists in general of a system of ordinary differential-algebraic equations for $\bm g = \bm g (t)$ function  of time only obtained by exploiting the electric-hydraulic analogy where the blood pressure and flow
rate are represented by voltage and current, respectively. For sake of simplicity we assume that $(\nabla\bm v )\bm n_o = \bm 0$ so that $|\bm g| = p$.
The downstream vasculature is modeled through a suitable combination of electrical elements: resistances $R$ accounting for the friction
effects, compliances $C$ for vessel deformation
and inductances $L$ for inertial effects leading to the so-called $RLC$ Windkessel models \cite{Shi:2011}. The numerical values for $R$, $C$ and $L$ need to be properly calibrated based on experimental measurements. 

\begin{figure}[h]
    \centering
{\includegraphics[width=8.0cm]{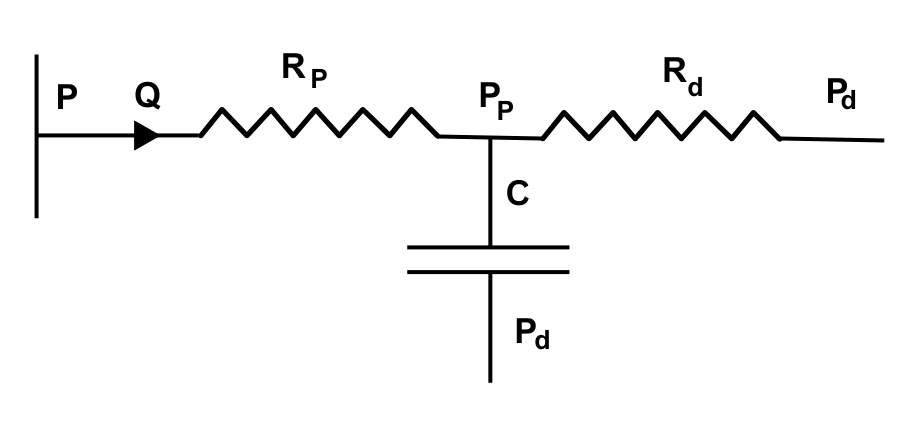} }
\caption{Thoracic aorta: $RCR$ Windkessel model for the generic outlet.}
\label{fig:RCR}
\end{figure}

For sake of simplicity, in this work we consider a three-element Windkessel $RCR$ model \cite{Westerhof:2008}. In short, it
consists of a proximal resistance $R_{p}$, a compliance $C$, and a distal resistance $R_{d}$, for each outlet: see Fig. \ref{fig:RCR}. 

The downstream pressure $p$ is computed by solving the following system:

\begin{equation}
   \left \{
   \begin{alignedat}{3}
    C \frac{dp_p}{dt} + \frac{p_p - p_d}{R_d}  & = Q & \quad & \mbox{ on } \quad \Gamma_{o} \times (0, T],\\
    p - p_p & = R_p Q & \quad & \mbox{ on } \quad \Gamma_{o} \times (0, T],
   \end{alignedat}
   \right .
   \label{eq:Windkessel-FOM}
\end{equation}
where $p_p$ is the proximal pressure, $p_d$ is the distal pressure (assumed to be, as often in literature \cite{Nichols:2022}, null, as it serves as a reference value) and $Q$ is the flow rate through the outlet section.


\subsection{Optimal control problem}\label{sec:opt}

In this approach, the normal stress $\bm g$ is computed by solving an optimal control problem
where the objective is to minimize the gap between computational results and clinical data.

From a formal viewpoint, an optimal control problem includes three elements: an objective functional to optimize, a control variable to choose in order to minimize the objective functional and a fluid flow model which represents a set of constraints for the optimization step \cite{Bewley:2001,Hak:2003,Gunzburger:2003,Quarteroni:2014}. For sake of simplicity, we consider the steady-state incompressible NSE as fluid flow model, i.e. the system \eqref{tN-Steady} with $\partial_t \bm v = \bm 0$, endowed with boundary conditions \eqref{tN-Steady-bc}. On the other hand,   $\bm g $ is the control variable and the objective functional is defined as follows: 
\begin{equation}
    \mathcal{I} (\bm v, \bm g) = \frac{1}{2} \int_{\Omega} |\bm v - \bm v_{\text{m}}|^2 \, d\Omega + \frac{\alpha}{2} \int_{\Gamma_o} |\bm g|^2 \, d\Gamma,
    \label{func}
\end{equation}
where $\bm v_{\text{m}}$ is the clinically observed velocity.  
Here, $\alpha > 0$ is a penalization parameter. The optimal control problem is to find $(\bm v, p, \bm g)$ 
such that the functional \eqref{func} is minimized subject to the constraint \eqref{tN-Steady}-\eqref{tN-Steady-bc} with $\partial_t \bm v = \bm 0$.

In order to solve this non-linear optimal control problem,  we adopt the adjoint-based Lagrangian method. In short the optimal control problem is recast in terms of an unconstrained minimization problem, whose solution corresponds to the minimum of a properly defined Lagrangian functional. The optimal solution is the one where all the derivatives of the Lagrangian functional vanish. The Lagrangian formulation requires the introduction of the so-called adjoint variables governed by the following equations 

\begin{equation}
	\begin{cases} 
     
- \nu \Delta \bm{w} - (\bm{v} \cdot \nabla) \bm{w} + \left(\nabla \bm{v}\right)^T \bm{w} - \nabla q = \bm{v} - \bm{v_{\text{m}}}  & \quad \text{in} \quad \Omega, \\
	\nabla \cdot \bm{w} = 0 & \quad \text{in} \quad \Omega, 
	\end{cases} \label{adj-Steady}
\end{equation}
where $\bm w$ is the adjoint velocity and $q$ is the adjoint pressure, endowed with the following boundary conditions:

\begin{equation}
    \begin{cases}
     \bm w = \bm 0  & \quad \text{on} \quad \Gamma_{i} \cup \Gamma_{w}, \\
     (\nabla\bm w )\bm n = \bm 0   & \quad \text{on} \quad \Gamma_{o} , \\
    \end{cases} \label{tN-Steady-bc2}
\end{equation}
while the optimality condition reads as follows
\begin{equation}
\alpha \bm g + \bm w = \bm 0 \quad 
\text{on} \quad \Gamma_o. 
\label{eq:adj}
\end{equation}

For further details, the reader is referred, e.g., to \cite{Gunzburger:2003,Quarteroni:2009}.

\section{Numerical results}\label{numerical_results}
In this section, we present some numerical results about two patient-specific cases, the former related to the numerical simulation of blood flow patterns in a thoracic aorta \cite{Girfoglio:2021,Girfoglio:2020}, the latter in a coronary artery system \cite{Zakia:2021,Balzotti:2022,Siena:2023}. In the first application, outlet boundary conditions are enforced using a lumped parameter model whereas in the second one an optimal control approach.

\subsection{Thoracic aorta}
\begin{figure}[h]
    \centering
{\includegraphics[width=4.0cm]{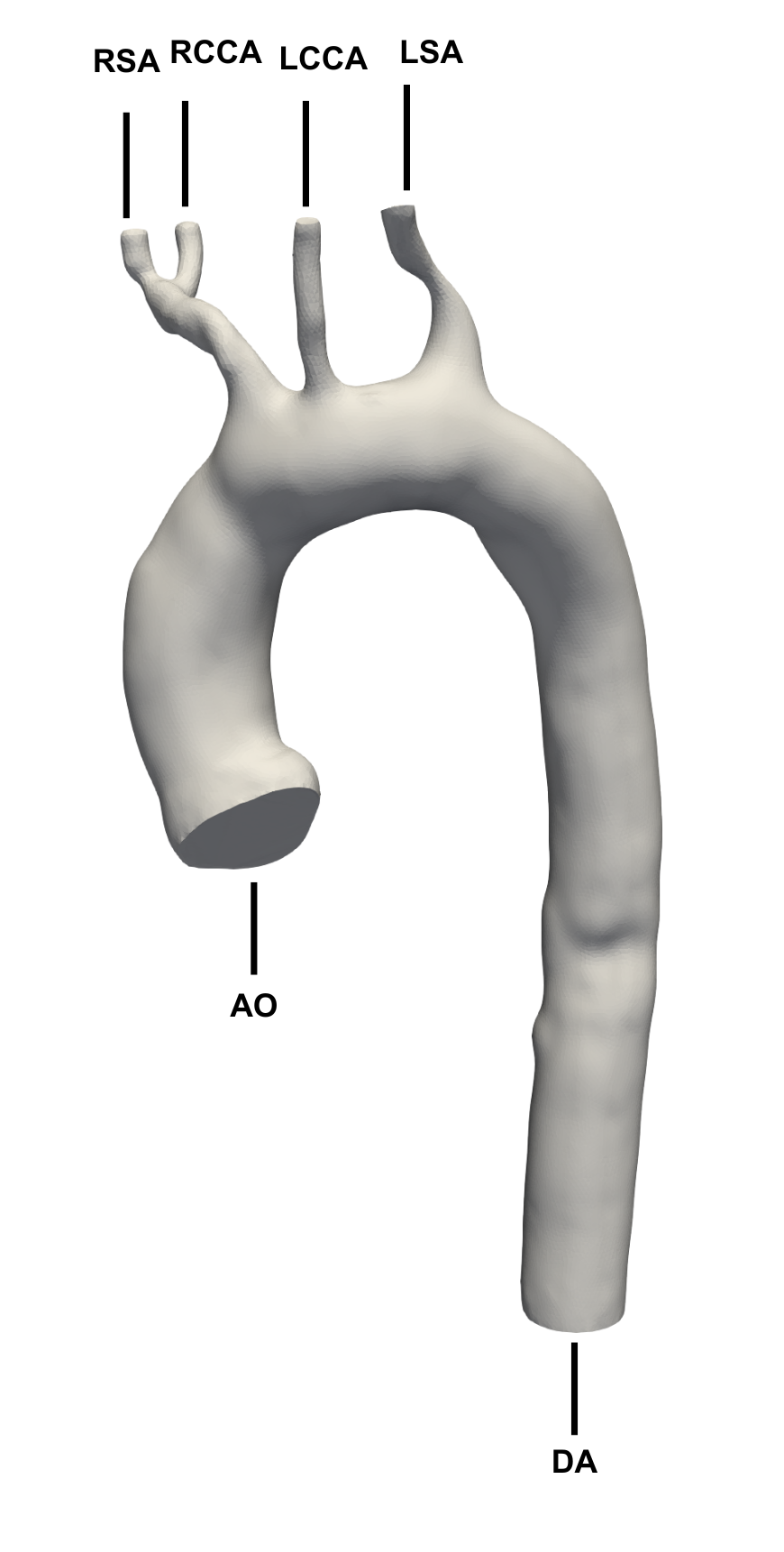} }
\caption{Thoracic aorta: sketch of the geometry.}
\label{fig:thoracic_aorta}
\end{figure}

The model analyzed includes  the ascending aorta (DA), the brachiocephalic artery (BCA), the right subclavian artery (RSA), the right common carotid artery (RCCA), the left common carotid artery (LCCA), the left subclavian artery (LSA) and the descending aorta (DA), as shown in Fig. \ref{fig:thoracic_aorta}.
\begin{figure}[h]
    \centering
{\includegraphics[width=10cm]{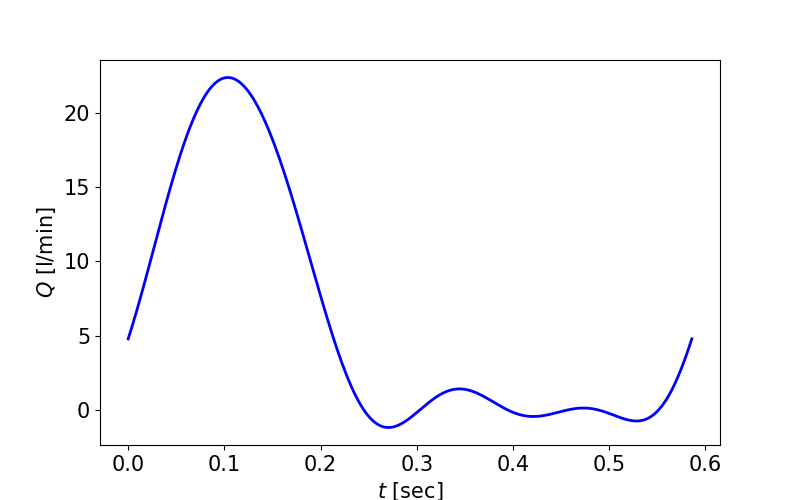} }
\caption{Thoracic aorta: inlet flow rate.}
\label{fig:inflow}
\end{figure}
We consider a mesh with 228296 tetrahedral elements. 
\begin{table}[h]
    \centering
    \caption{Thoracic aorta: experimental data obtained by the RHC and ECHO tests. SBP = systolic blood pressure, DBP = diastolic blood pressure, MAP = mean arterial pressure, CO = average cardiac flow rate, SV = stroke volume.}
    \label{tab:pre_surgery}
    \begin{tabular}{ccccc}
        \toprule
        SBP [mmHg] & DBP [mmHg] & MAP [mmHg] & CO [l/min] & SV [ml] \\
        \midrule
        108 & 66 & 78 & 5.63 & 55 \\
        \bottomrule
    \end{tabular}
\end{table}
 Experimental measurements have been obtained by  Right Heart Catheterization (RHC) and Echocardiography (ECHO) tests and are reported in Table \ref{tab:pre_surgery} \cite{Girfoglio:2021,Girfoglio:2020}. 
 
Concerning the inlet boundary condition, a realistic flow rate $Q$ was enforced on AO section: see Fig. \ref{fig:inflow}. The amplitude of the flow waveform has been set according to 
\begin{equation}
\text{CO} = \dfrac{1}{T} \int_0^T Q dt.
\end{equation}
The period of the cardiac cycle $T$ is estimated as
\begin{equation}
T = \dfrac{\text{SV}}{\text{CO}} = 0.586 \hspace{0.1cm} \text{s}.
\end{equation}

Outflow boundary conditions were applied at each outlet of the model (namely RSA, RCCA, LCCA, LSA and DA) by using a three-element Windkessel $RCR$ model \eqref{eq:Windkessel-FOM}.  The total resistance at each 
outlet $k$, $R_k = R_{p,k} + R_{d,k}$, can be evaluated following the rules for a 
parallel circuit:
\begin{equation}
R_k = \text{SVR} \dfrac{\sum_k A_k}{A_k},
\end{equation}
where $A_k$ is the cross-sectional area of outlet $k$ and $\text{SVR}$ is  the systemic vascular resistance estimated as follows 
\begin{equation}
SVR = \dfrac{\text{MAP}}{\text{CO}} = 1105 \hspace{0.1cm} \text{g} \cdot \text{s}^{-1} \cdot \text{cm}^{-4}.
\end{equation}
For each outlet $k$, following \cite{Laskey:1990} we assumed 
$R_{p,k}/R_k = 0.056$. On the other hand, the total aortic compliance $C$ can be estimated as 
follows \cite{Bulpitt:1999}: 
\begin{equation}
C = \dfrac{\text{SBP} - \text{DBP}}{\text{SV}} = 9.85 \times 10^{-4} \hspace{0.1cm} \text{g}^{-1} \cdot \text{s}^2 \cdot \text{cm}^4.
\end{equation}
The compliance $C_k$, related to outlet $k$, can be 
evaluated following the rules for a parallel circuit: 
\begin{equation}
C_k = C \dfrac{A_k}{\sum A_k}.
\end{equation}
Table \ref{tab:windkessel} reports the values of 
Windkessel coefficients.


\begin{table}[h]
    \centering
    \caption{Thoracic aorta: Windkessel coefficients for each outlet $k$.}
    \label{tab:windkessel}
    \begin{tabular}{lccc}
        \hline
        $k$ & $R_{p,k}$ [$\text{g} \cdot \text{s}^{-1} \cdot \text{cm}^{-4}$] & $R_{d,k}$ [$\text{g} \cdot \text{s}^{-1} \cdot \text{cm}^{-4}$] & $C_k$ [$\text{g}^{-1} \cdot \text{s}^2 \cdot \text{cm}^4$] \\
        \hline
        RSA & $1.84 \times 10^3$ & $3.11 \times 10^4$ & $3.26 \times 10^{-5}$ \\
        RCCA & $1.23 \times 10^3$ & $2.07 \times 10^4$ & $5.16 \times 10^{-5}$ \\
        LCCA & $1.78 \times 10^3$ & $3.01 \times 10^4$ & $3.52 \times 10^{-5}$ \\
        LSA & $7.09 \times 10^2$ & $1.19 \times 10^4$ & $9.35 \times 10^{-5}$ \\
        DA & $7.80 \times 10^1$ & $1.31 \times 10^3$ & $7.72 \times 10^{-4}$ \\
        \hline
    \end{tabular}
\end{table}


For the space discretization we adopt a Finite Volume  approximation whilst for the time discretization we employ an implicit Euler method.  All the computational results presented in this article have been performed with OpenFOAM \cite{Weller:1988}, an open source C++ library widely used by commercial and academic organizations. 
The time step used for the simulation is 10$^{-3}$ s. 
For further details about the computational setup the reader is referred to \cite{Girfoglio:2021,Girfoglio:2020}.

The comparison between computational and experimental data is 
carried out in terms of $\text{SBP}$, $\text{DBP}$ and $\text{MAP}$. Computational estimates 
of such quantities are obtained as: 
\begin{align}
    \text{SBP} &= \max_{t \in [0,T]} p_{\text{avg}}, \label{eq:SBP} \\
    \text{DBP} &= \min_{t \in [0,T]} p_{\text{avg}}, \label{eq:DBP} \\
    \text{MAP} &= \frac{1}{T} \int_{0}^{T} p_{\text{avg}} \, dt, \label{eq:MAP}
\end{align}
where $p_{avg}$ is the space-averaged pressure
\begin{equation}
    p_{\text{avg}} = \frac{1}{\Omega} \int_{\Omega} p \, d\Omega.
\end{equation}

Table \ref{tab:exp_comp} 
reports numerical results and experimental data. We observe that the 
error is very small for $DBP$ and $MAP$: about 4\% and 2.4\%, respectively. On the other hand, the error for $SBP$ is greater, around 11.7\%, but still satisfactory. 
\begin{table}[h]
    \centering
    \caption{Thoracic aorta: comparison between numerical outcomes and experimental data in mmHg.}
    \label{tab:comparison}
    \begin{tabular}{cccccc}
        \toprule
        SBP (exp) & SBP (num) & DBP (exp) & DBP (num) & MAP (exp) & MAP (num) \\
        \midrule
        108 & 95.4 & 66 & 63.4 & 78 & 79.9 \\
        \bottomrule
    \end{tabular}\label{tab:exp_comp}
\end{table}

\begin{figure}
\centering
    {\includegraphics[width=11cm]{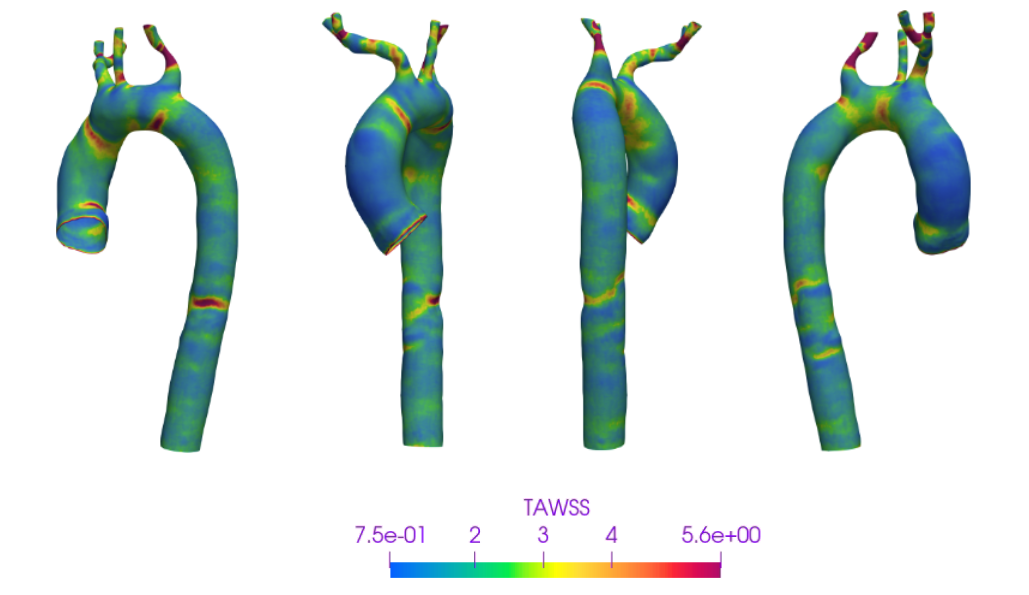} }
    \caption{Thoracic aorta: TAWSS (in Pa) magnitude distribution on the aortic 
wall.}
    \label{fig:WSS}
\end{figure}

Finally we show in Fig. \ref{fig:WSS} the distribution of the time-averaged wall shear stress $\text{TAWSS}$ defined as 
\begin{equation}
    \text{TAWSS} = \frac{1}{T} \int_{0}^{T} \left(\boldsymbol{\tau}(\bm v) \cdot \bm{n}_w\right)  dt \quad \text{on } \Gamma_w, 
\end{equation}
where $\boldsymbol{\tau} (\bm v) = \nu \nabla \bm v$ and  $\bm{n}_w$ is the outward unit normal vector to \( \Gamma_w\).  

There are no available clinical data for the validation of this hemodynamic indicator, so we limit to give some basic insights at the aim to justify the patterns obtained.  
TAWSS peak values are localized in portions of the domain where either the cross section narrows or large curvatures are present. On the other hand, regions characterized by lower TAWSS correspond to cross-sectional enlargements of the vasculature. This trend is in perfect agreement with the well-known behaviour of a fully developed flow in a cylindric pipe.

\subsection{Coronary artery system}

\begin{figure}[h]
    \centering
    \includegraphics[width=0.5\textwidth]{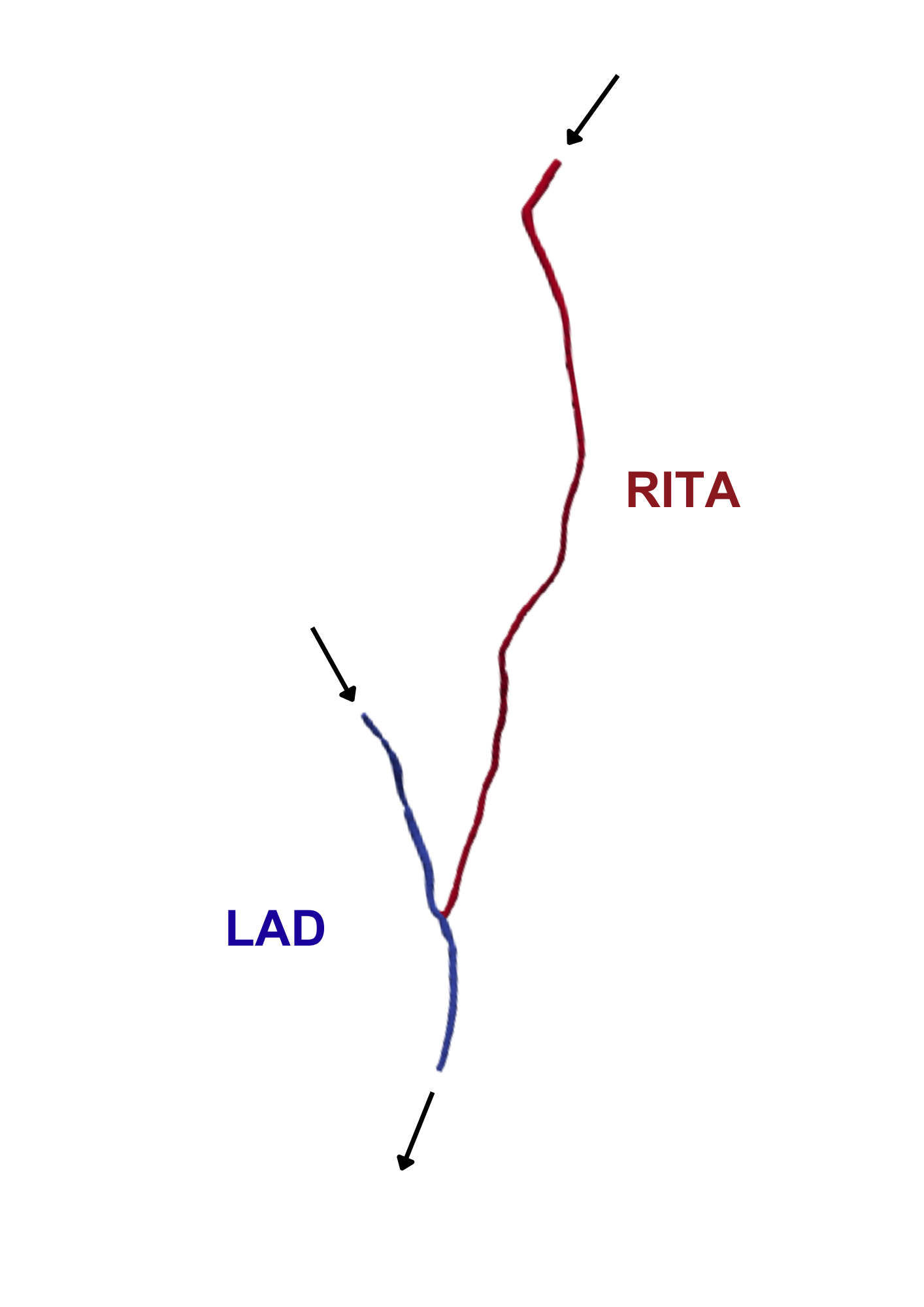}
    \caption{Coronary artery system: sketch of the geometry. 
    The arrows indicate the flow direction.}
    \label{fig:geometry}
\end{figure}

The geometry analyzed features a coronary artery bypass graft (CABG) performed with the right
internal thoracic artery (RITA) on the left anterior descending artery (LAD). The geometry is shown in Figure \ref{fig:geometry}. We have two inlets, the upper cross sections of RITA and LAD, and an outlet, The lower cross section of LAD. 
 We consider a mesh with 42354 tetrahedral elements.

To solve the nonlinear optimal control problem discussed in Sec. \ref{sec:opt}, we adopt an \textit{optimize-then-discretize} approach,  i.e. at first we derive optimality
 conditions as system \eqref{tN-Steady}-\eqref{tN-Steady-bc}-\eqref{adj-Steady}-\eqref{tN-Steady-bc2}-\eqref{eq:adj} (with $\partial_t \bm v = \bm 0$) and then we discretize it. 
The discretization is performed by the Finite Element method  employed in the Python libraries FEniCS and multiphenics \cite{Alnaes:2015,Logg:2012}.  
To ensure the uniqueness of the solution, inf-sup stable $\mathbb{P}_2-\mathbb{P}_1$ elements are used for velocity and pressure, respectively. We set $\alpha = 10^{-2}$. 
See, e.g., \cite{Gunzburger:2003,Balzotti:2022,Schulz:2009,Ballarin:2020, Ballarin:2015} 
for further
details.

Let $\bm t_c$ be the tangent vector to the vessel centerline, 
$R$ the maximum vessel radius, and $r$ the distance from mesh nodes to the centerline. The desired blood flow velocity is defined as:
\[
\bm v_{\mathrm{m}} = v_{\text{const}} \Big(1-\frac{r^2}{R^2}\Big)\bm t_c,
\]
where $v_{\text{const}} = 350\,\mathrm{mm/s}$ is the desired velocity magnitude. 

At inlets, we prescribe:
\begin{equation}\label{eq:vinlet}
    \bm v_{i} = \frac{\nu Re}{R_{i}}\Big(1-\frac{r^2}{R_{i}^2}\Big)\bm n_{i},
\end{equation}
where $\nu = 3.6\,\mathrm{mm^2/s}$, 
\(R_{i}\) is the maximum inlet radius and \(\bm n_{i}\) is the outward unit normal vector to $\Gamma_i$. 

\begin{figure}
    \centering
    \begin{minipage}{0.45\textwidth}
        \centering
        \subfloat[\centering Pressure]{{\includegraphics[width=3.5cm]{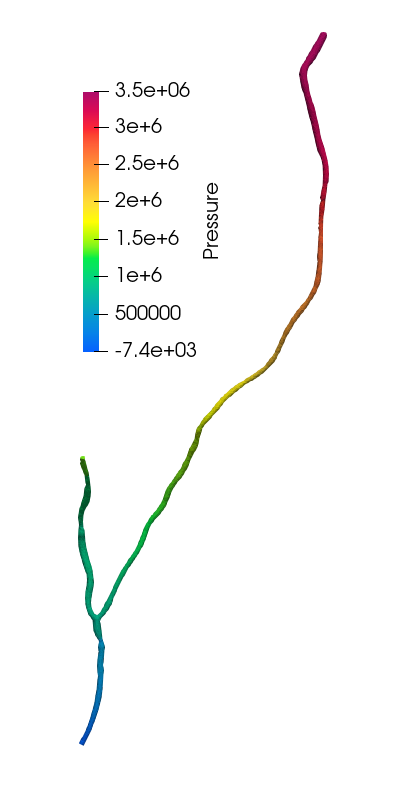} }}
    \end{minipage}
    \hfill
    \begin{minipage}{0.45\textwidth}
        \centering
        \subfloat[\centering Velocity magnitude]{{\includegraphics[width=6.8cm, height=14cm, keepaspectratio]{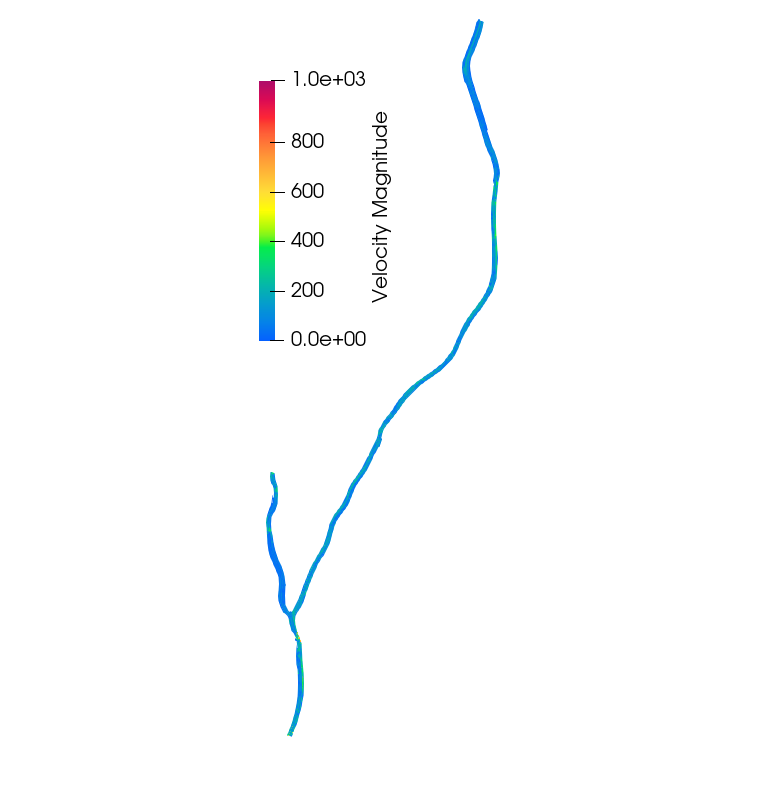} }}
    \end{minipage}
    \caption{Coronary artery system: pressure (mm$^2$/s$^2$) and velocity magnitude (mm/s$^2$) distributions.}
\label{fig:velocity_pressure}
\end{figure}

\begin{figure}[h]
    \centering
    \includegraphics[width=0.65\textwidth]{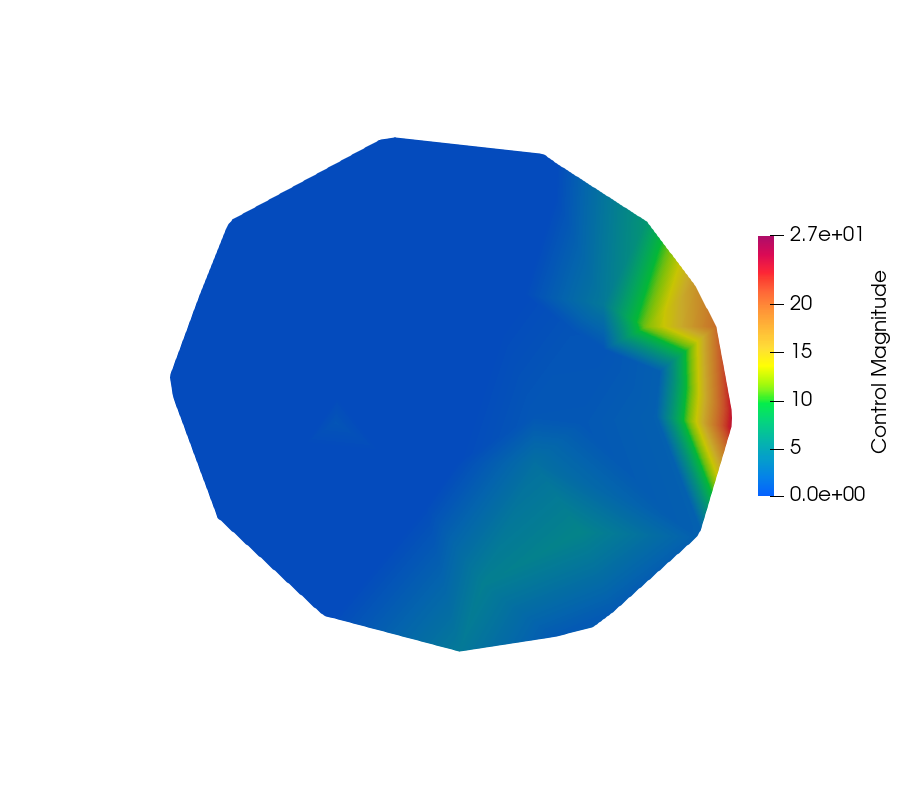}
    \caption{Coronary artery system: control variable magnitude (mm$^2$/s$^2$) distribution at the outflow boundary,}  
    \label{fig:control}
\end{figure}

Figure \ref{fig:velocity_pressure} shows the pressure and the velocity for $Re = 75$.
Moreover, Figure \ref{fig:control} shows the control variable at the outflow boundary. 
We observe that the distribution of the control variable is strongly not homogeneous. 
This suggests that the optimal control approach could potentially provide more accurate results with respect to the lumped element model, the latter one assuming a uniform distribution of the pressure/shear stress on the outlet section.  

\section{Conclusion}\label{conclusion}

In this chapter we have presented 
two different ways for the estimation of proper outlet boundary conditions in the numerical simulations of the cardiovascular system. The first one is related to a lumped parameter model describing the outlet physiological conditions by means of a suitable combination of electrical elements. The second one instead involves an optimal control
problem where the mismatch between numerical solution and clinical data is minimized by controlling the normal stress at the outlet section. Two patient-specific cases have been investigated, the one regarding the hemodynamics in the thoracic aorta where the outlet boundary conditions have been treated by a lumped element model, the other one concerning the hemodynamics in a coronary artery system where the tuning of the outlet boundary conditions has been performed by the optimal control approach. Both procedures have been able to provide promising results. We highlight that the optimal control approach has the significant advantage to take into account the space variability of the normal stress across the outlet section. 

\section*{Aknowledgments}
We acknowledge INdAM-GNCS projects, the PRIN 2022 Project ``Machine
 learning for fluid-structure interaction in cardiovascular problems: efficient solutions, model reduction, inverse problem" and the consortium iNEST (Interconnected North-East Innovation Ecosystem), Piano Nazionale di Ripresa
 e Resilienza (PNRR) supported by the European Union's NextGenerationEU program.


\begin{thebibliography}{99.}%
%
%
\bibitem{Siena:2024}
P. Siena, P. C. Africa, M. Girfoglio, and G. Rozza. Chapter Six - On the accuracy and efficiency of reduced order models: Towards real-world applications. \textit{Advances in Applied Mechanics}, volume 59, pages 245-288, 2024.


\bibitem{Morbiducci:2010}
U. Morbiducci, D. Gallo, D. Massai, F. Consolo, R. Ponzini, L. Antiga, C. Bignardi, M. A. Deriu, A. Redaelli. Outflow conditions for image-based hemodynamic models of the carotid bifurcation: implications for indicators of abnormal flow. \textit{Journal of Biomechanical Engineering}, 132(9): 091005-1–091005-11, 2010.

\bibitem{Van:2011}
A. G. Van der Giessen, H. C. Groen, P. A. Doriot, P. J de Feyter, A. F. W van der Steen, F. N van de Vosse, J. J Wentzel, F. J. H Gijsen. The influence of boundary conditions on wall shear stress distribution in patients specific coronary trees. \textit{Journal of Biomechanics}, 44(6): 1089-1095, 2011.

\bibitem{Pedley:1980}
T. J. Pedley. The fluid mechanics of large blood vessels. \textit{Cambridge: Cambridge University Press}, 1980.

\bibitem{Sankaran:2012}
S. Sankaran, M. E. Moghadam, A. E. Kahn, E. E. Tseng, J. M. Guccione, A. L. Marsden. Patient-specific multiscale modeling of blood flow for coronary artery bypass graft surgery. \textit{Annals of Biomedical Engineering}, 40(10): 2228-2242, 2012.

\bibitem{Grinberg:2008}
L. Grinberg, G. E. Karniadakis. Outflow boundary conditions for arterial networks with multiple outlets. \textit{Annals of Biomedical Engineering}, 36(9): 1496-1514, 2008.

\bibitem{Zakia:2021}
Z. Zainib, F. Ballarin, S. Fremes, P. Triverio, L. Jim\'{e}nez-Juan, and G. Rozza. Reduced order methods for parametric optimal flow control in coronary bypass grafts, toward patient-specific data assimilation. \textit{International Journal for Numerical Methods in Biomedical Engineering}, 37(12):Paper No. e3367, 24, 2021.

\bibitem{Fevola:2021}
E. Fevola, F. Ballarin, L. Jim\'{e}nez-Juan, S. Fremes, S. Grivet-Talocia, G. Rozza, and 
P. Triverio. An optimal control approach to determine resistance-type boundary conditions from in-vivo data for cardiovascular simulations. \textit{International Journal for Numerical Methods in Biomedical Engineering}, 37(10):e3516, 2021.

\bibitem{Shi:2011}
Y. Shi, P. Lawford, and R. Hose. Review of zero-d and 1-d models of blood flow in the cardiovascular system. \textit{Biomedical engineering online}, 10:33, 2011.

\bibitem{Westerhof:2008}
N. Westerhof, J. W. Lankhaar, and B. Westerhof. The arterial Windkessel. \textit{Medical \& biological engineering \& computing}. 47:131–41, 2008.

\bibitem{Nichols:2022}
W. W. Nichols, M. O'Rourke, E. R. Edelman, and C. Vlachopoulos. McDonald’s blood flow in arteries: theoretical, experimental and clinical principles.  
\textit{CRC Press}, 2022.

\bibitem{Bewley:2001} T. R. Bewley. Flow control: new challenges for a new renaissance. \textit{Progress in Aerospace
sciences}, 37(1):21–58, 2001.

\bibitem{Hak:2003} 
M. Gad-el Hak, A. Pollard, and J. P. Bonnet. Flow control: fundamentals and practices, volume 53. \textit{Springer Science \& Business Media}, 2003. 

\bibitem{Gunzburger:2003} M. D. Gunzburger. \textit{Perspectives in flow control and optimization}, volume 5 of \textit{Advances
in Design and Control}, Society for Industrial and Applied Mathematics (SIAM), Philadelphia, PA, 2003. 


\bibitem{Quarteroni:2014}
A. Quarteroni. Numerical models for differential problems, volume 8 of MS\&A. \textit{Modeling, Simulation and Applications}, Springer, Milan, 2014.

\bibitem{Quarteroni:2009}
A. Quarteroni, S. Quarteroni. Numerical Models for Differential Problems. Vol 2. Milan: \textit{Springer}, 2009.

\bibitem{Girfoglio:2021}
M. Girfoglio, L. Scandurra, F. Ballarin, G. Infantino, F. Nicolo, A. Montalto, G. Rozza,
R. Scrofani, M. Comisso, and F. Musumeci. Non-intrusive data-driven rom framework for
hemodynamics problems. \textit{Acta mechanica sinica}, 37(7):1183–1191, 2021.

\bibitem{Girfoglio:2020}
M. Girfoglio, F. Ballarin, G. Infantino, F. Nicolo, A. Montalto, G. Rozza, R. Scrofani,  M. Comisso, and F. Musumeci. Non-intrusive PODI-ROM for patient-specific aortic blood flow
in presence of a LVAD device. \textit{Medical Engineering \& Physics}, 107, 103849, 2022.


\bibitem{Balzotti:2022}
C. Balzotti, P. Siena, M. Girfoglio, A. Quaini, and G. Rozza.
A data-driven reduced order method for parametric optimal blood flow control: Application to coronary bypass graft. In: \textit{Communications in Optimization Theory}, 26, 1–19, 2022.

\bibitem{Siena:2023}
P. Siena, M. Girfoglio, F. Ballarin, G. Razza. Data-Driven Reduced Order Modelling for Patient-Specific Hemodynamics of Coronary Artery Bypass Grafts with Physical and Geometrical Parameters. \textit{Journal of Scientific Computing}, 94, 38, 2023. 


\bibitem{Laskey:1990}
W. Laskey, H. Parker, V. Ferrari, W. Kussmaul, A. Noordergraaf. Estimation of total systemic arterial compliance in humans. \textit{Journal of Applied Physiology}, 1990;69: 112–9, 1990.

\bibitem{Bulpitt:1999}
C. Bulpitt, J. Cameron, C. Rajkumar, S. Armstrong, M. Connor, J. Joshi, D. Lyons, O. Moioli, P. Nihoyannopoulos. The effect of age on vascular compliance in man: Which are the appropriate measures? \textit{Journal of Human Hypertension} 1999;13: 753–8.

\bibitem{Weller:1988}
H. G. Weller, G. Tabor, H. Jasak; C. Fureby. A tensorial approach to computational continuum mechanics using object-oriented techniques.\textit{ Computers in Physics}, 12, 620–631, 1998.

\bibitem{Alnaes:2015}
M. Aln\ae{s}, J. Blechta, J. Hake, A. Johansson, B. Kehlet, A. Logg, C. Richardson, J. Ring, M. E. Rognes, and G. N. Wells. The fenics project version 1.5. \textit{Archive of Numerical
Software}, 3(100), 2015.

\bibitem{Logg:2012}
A. Logg, K. A. Mardal, and G. N. Wells, editors. Automated solution of differential equations by the finite element method, volume 84 of \textit{Lecture Notes in Computational Science and Engineering}, Springer, Heidelberg, 2012. The FEniCS book.


\bibitem{Schulz:2009}
V. Schulz and I. Gherman. One-shot methods for aerodynamic shape optimization. \textit{In MEGADESIGN and MegaOpt-German Initiatives for Aerodynamic Simulation and Optimization in Aircraft Design}, pages 207–220. Springer, 2009.

\bibitem{Ballarin:2020}
S. Ali, F. Ballarin, and G. Rozza. Stabilized reduced basis methods for parametrized steady Stokes and Navier-Stokes equations. \textit{Computers \& Mathematics with Applications}, An International Journal, 80(11):2399–2416, 2020.

\bibitem{Ballarin:2015}
F. Ballarin, A. Manzoni, A. Quarteroni, and G. Rozza. Supremizer stabilization of POD-Galerkin approximation of parametrized steady incompressible Navier-Stokes equations. \textit{International Journal for Numerical Methods in Engineering}, 102(5):1136–1161, 2015.


\end{thebibliography}
\end{document}